\documentclass[11pt,epsf,letterpaper]{article}%
\usepackage[usenames,dvipsnames,svgnames,table]{xcolor}
\usepackage{color}
\usepackage{amsmath}
\usepackage{amsfonts}
\usepackage{verbatim}
\usepackage{amssymb}
\usepackage{graphicx}
\usepackage{epstopdf}
\usepackage{mathrsfs}%
\providecommand{\U}[1]{\protect\rule{.1in}{.1in}}
\begin{document}

\title{Exact asymptotically flat charged hairy black holes \\with a dilaton potential}
\author{$^{(1)}$Andr\'{e}s Anabal\'{o}n, $^{(2)}$Dumitru Astefanesei and $^{(3,4)}%
$Robert Mann.\\\textit{$^{(1)}$Departamento de Ciencias, Facultad de Artes Liberales y}\\\textit{Facultad de Ingenier\'{\i}a y Ciencias, Universidad Adolfo
Ib\'{a}\~{n}ez, Vi\~{n}a del Mar, Chile.}\\\textit{$^{(2)}$Instituto de F\'\i sica, Pontificia Universidad Cat\'olica de
Valpara\'\i so,} \\\textit{ Casilla 4059, Valpara\'{\i}so, Chile.}\\\textit{$^{(3)}$ Department of Physics and Astronomy, University of Waterloo,}\\\textit{Waterloo, Ontario, Canada N2L 3G1.}\\\textit{$^{(4)}$ Perimeter Institute, 31 Caroline Street North Waterloo,
Ontario Canada N2L 2Y5.}}
\maketitle

\begin{abstract}
We find broad classes of exact 4-dimensional asymptotically flat black hole
solutions in Einstein-Maxwell theories with a non-minimally coupled dilaton
and its non-trivial potential. We consider a few interesting limits, in
particular, a regular generalization of the dilatonic Reissner-Nordstr{\"o}m
solution and, also, smooth deformations of supersymmetric black holes. Further
examples are provided for more general dilaton potentials. We discuss the
thermodynamical properties and show that the first law is satisfied. In the
non-extremal case the entropy depends, as expected, on the asymptotic value of
the dilaton. In the extremal limit, the entropy is determined purely in terms
of charges and is independent of the asymptotic value of the dilaton. The
attractor mechanism can be used as a criterion for the existence of the
regular solutions. Since there is a `competition' between the effective
potential and dilaton potential, we also obtain regular extremal black hole
solutions with just one U(1) gauge field turned on.

\end{abstract}

\section{Introduction.}

No-hair theorems \cite{Israel:1967wq} apply to stationary asymptotically flat
black holes in theories of gravity coupled to scalar fields with an
interaction potential that is convex \cite{Bekenstein:1995un} or positive
semidefinite \cite{Sudarsky:1995zg}. It is perhaps unsurprising that if one or
more of these assumptions are relaxed, then the theorem does not hold. As is
well known, the scalar potential of compactified supergravity\footnote{The
effective potential typically has negative regions but supersymmetry ensures
the total energy is always positive.} (see, \cite{Hertog:2006rr} for a
discussion in the context of no-hair theorems) does not necessary satisfy the
constraints mentioned above and one can ask if the no-hair theorems could be
evaded for some potentials of scalar fields. It is by now well known that, in
asymptotically anti de Sitter (AdS) spacetimes, there exist indeed (numerical,
as well as exact) scalar-hairy black holes for some scalar field potentials
(see, e.g., \cite{Torii:2001pg, Sudarsky:2002mk, Anabalon:2012ta}).

In AdS, the existence of hairy black holes is related to the fact that the
asymptotic value of the scalar field potential can be reabsorbed in an
effective cosmological constant that differs from the true cosmological
constant of the theory. Interestingly, it has been suggested in
\cite{Sudarsky:2002mk} that the effective cosmological constant could vanish
if the parameters of the scalar field potential are properly adjusted and
numerical evidence for the existence of the hairy asymptotically flat black
holes was presented. Therefore, by fine tuning the cosmological constant or
the scalar field potential so that the effective cosmological vanishes, one
can obtain scalar field hair for asymptotically flat black holes. The scalar
field potentials considered in \cite{Sudarsky:2002mk} are not positive
semidefinite and the weak energy condition is violated.

Behind the no-hair theorems there is a simple physical picture: the matter
fields left in the exterior after the gravitational collapse would eventually
be `swallowed' by the black hole itself or be radiated away to
infinity.\footnote{The limited information about the matter that collapsed is
reflected \textit{only} in the number of conserved charges that are conserved
surface integrals in the asymptotically flat region. `Hair' does not refer to
those fields, which are associated with conserved charges.} Since the `hair'
is located outside the horizon, one could ask how is it possible that matter
can hover in a strong gravitational field without collapsing completely?
Intuitively, the answer could be that this is possible if the internal
pressure is sufficiently large. However, this kind of intuition works just
near the horizon and it should be clear that the the non-linear
(self-interacting) character of the matter fields is, in fact, at the basis of
the existence of hairy black holes. A nice heuristic picture was presented in
\cite{Nunez:1996xv}: the self-interaction of matter fields (together with the
gravitational interaction) is responsible for the fact that the near-horizon
hair does not collapse into the black hole while the far-region hair does not
escape to infinity. Therefore, the hair should extend some way out from the
event horizon so that the degrees of freedom near the horizon are bound
together with the ones that tend to be radiated away at infinity.

In this paper, we present exact non-extremal asymptotically flat
charged hairy black holes with a dilaton potential and investigate
their thermodynamic properties.\footnote{Exact charged hairy black
holes in Anti-de Sitter spacetime will be presented elsewhere
\cite{noi}. Exact hairy black hole solutions in Anti-de Sitter when
the gauge fields are turned off were presented in
\cite{Anabalon:2012ta} and their holographic properties were
investigated in \cite{Salvio:2012at, Anabalon:2012ta}.} In
particular, we are able to recover some of the solutions of
\cite{Gibbons:1987ps} when the dilaton potential vanishes. In
\cite{Gibbons:1987ps}, the authors found the black hole solutions by
rewriting the equations of motion as Toda equations. However, when
there is a non-trivial dilaton potential, this method of solving the
equations of motion is not very useful. Instead, we use the metric
ansatz of \cite{Anabalon:2012ta} to obtain charged hairy black hole
solutions, which are asymptotically flat.

In the non-extremal case, the near horizon data depend on the value of the
dilaton at the horizon. In the extremal case, due to the attractor mechanism
\cite{Ferrara:1995ih}, the near horizon data are completely fixed by the
electric and magnetic charges and so the attractor mechanism acts as a no-hair
theorem \cite{Israel:1967wq} for the extremal black holes
\cite{Astefanesei:2007vh}. The attractor mechanism is a useful tool in
investigating the existence of the regular extremal black hole solutions.

The remainder of the paper is organized as follows: In Section \ref{gensol},
we construct the hairy black hole solutions in flat space. First, we obtain a
general solution and then we discuss some concrete examples which are related
to some known solutions when the dilaton potential vanishes. In Section
\ref{hairyy} we explicitly compute the thermodynamic quantities and show that
the first law is satisfied. In Section \ref{attractor} we use the attractor
equations to investigate the extremal limit. We show that at zero temperature,
contrary to the case for which the dilaton potential vanishes, there exist
regular solutions with only one gauge field (electric or magnetic) turned on.
Finally, we end with a detailed discussion of our results in Section
\ref{discuss} including the energy conditions. In the appendices we present
more general solutions and detail our progress in looking for analytic solutions.

\bigskip

\bigskip

\bigskip

\section{Non-extremal solutions.}

\label{gensol}

In this section, we construct static asymptotically flat non-extremal black
holes for a model with one scalar field (dilaton) non-minimally coupled to a
gauge field and a non-trivial dilaton potential. For an exponential coupling,
the equations of motion can be solved exactly with a special metric ansatz and
the dilaton potential can be explicitly obtained. Interestingly enough, in
this way we obtain analytic generalizations of some well known solutions, in
particular the dilatonic Reissner-Nordstr{\"{o}}m and other solutions for
which the extremal limits are supersymmetric. The action we are interested in
is
\begin{equation}
I[g_{\mu\nu},A_{\mu},\phi]=\frac{1}{2\kappa}\int d^{4}x\sqrt{-g}\left[
R-\frac{1}{4}e^{\gamma\phi}F^{2}-\frac{1}{2}\partial_{\mu}\phi\partial^{\mu
}\phi-V(\phi)\right]
\end{equation}
where the gauge coupling and potential are functions of the dilaton and we use
the convention $\kappa=8\pi G_{N}$. Since we set $c=1=\hbar$, $\left[
\kappa\right]  =M_{P}^{-2}$ where $M_{P}$ is the reduced Planck mass. The
equations of motion for the gauge field, dilaton, and metric are
\begin{equation}
\nabla_{\mu}\left(  e^{\gamma\phi}F^{\mu\nu}\right)  =0
\end{equation}%
\begin{equation}
\frac{1}{\sqrt{-g}}\partial_{\mu}\left(  \sqrt{-g}g^{\mu\nu}\partial_{\nu}%
\phi\right)  -\frac{\partial V}{\partial\phi}-\frac{1}{4}\gamma e^{\gamma\phi
}F^{2}=0\label{dil}%
\end{equation}%
\begin{equation}
R_{\mu\nu}-\frac{1}{2}g_{\mu\nu}R=\frac{1}{2}\left[  T_{\mu\nu}^{\phi}%
+T_{\mu\nu}^{EM}\right]
\end{equation}
where the stress tensors of the matter fields are
\begin{equation}
T_{\mu\nu}^{\phi}=\partial_{\mu}\phi\partial_{\nu}\phi-g_{\mu\nu}\left[
\frac{1}{2}\left(  \partial\phi\right)  ^{2}+V(\phi)\right]
\,\,\,\,,\,\,\,\,\,T_{\mu\nu}^{EM}=e^{\gamma\phi}\left(  F_{\mu\alpha}F_{\nu
}^{\cdot\alpha}-\frac{1}{4}g_{\mu\nu}F^{2}\right)
\end{equation}
It is still very difficult to find exact solutions to the system of
coupled differential equations above when the dilaton potential does
not vanish, even for simple potentials \cite{Garfinkle:1990qj,
Gregory:1992kr, Charmousis:2009xr}. For simplifying our analysis, we
use the following spherically symmetric ansatz for the metric:
\begin{equation}
ds^{2}=\Omega(x)\left[  -f(x)dt^{2}+\frac{\eta^{2}dx^{2}}{f(x)}+d\theta
^{2}+\sin^{2}{\theta}d\varphi^{2}\right]  \label{Ansatz}%
\end{equation}
where the parameter $\eta$ was introduced to obtain a dimensionless radial
coordinate $x$ and $\Omega(x)$ is the conformal factor. Written in this way,
it is clear that $\eta$ is going to be the only constant of integration of
gravitational origin in the solution (it does not appear in the action) and so
it should be related to the mass of the solution. This is the most general
static asymptotically flat solution and so it is characterized by only two
unknown functions.

The equation of motion for the gauge field and Bianchi identity can be solved
via the following ansatz for the field strength:%
\begin{equation}
F=Qe^{-\gamma\phi}dx\wedge dt+P\sin{\theta d\varphi}\wedge d\theta\label{GF}%
\end{equation}

Using the observation that the only terms involving the dilaton potential in
the Einstein equations are of the form $\delta^{\mu}_{\nu}V(\phi)$, we can use
combinations for which the dilaton potential cancels out. We then explicitly
obtain a differential equation that involves only the dilaton and conformal
factor $\Omega(x)$:%
\begin{equation}
E^{t}_{t}-E^{x}_{x}=0\,\,\,\, \rightarrow\,\,\,\,\, \phi^{\prime2}%
=\frac{-2\Omega^{\prime\prime}\Omega+3\Omega^{\prime2}}{\Omega^{2}}
\label{scalar}%
\end{equation}
where $\Omega^{\prime}$ is the derivative with respect to the coordinate $x$
and $E_{\mu}^{\nu}$ are the Einstein field equations.

For some special conformal factors $\Omega(x)$, the equation above can be
exactly solved and, in this way, we are able to obtain the analytic solutions
presented below. We would also like to point out that one can get the `hairy'
Reissner-Nordstr{\"o}m solution by taking the limit $\gamma=0$.

The strategy for finding solutions is to use the same conformal
factor as in \cite{Anabalon:2012ta}
\begin{equation}
\Omega(x)=\frac{\nu^{2}x^{\nu-1}}{\eta^{2}\left(  x^{\nu}-1\right)  ^{2}}
\label{CF}%
\end{equation}
then find the scalar field from (\ref{scalar}). We would like to emphasize
that the new parameter $\nu$ labels different solutions, for example $\nu=1$
and $\gamma=0$ corresponds to Reissner-Nordstr{\"{o}}m black hole, and for
$\nu>1$ we obtain `hairy' solutions.

From (\ref{scalar}) it follows immediately that%

\begin{equation}
\phi(x)=l_{\nu}^{-1}\ln(x)+\phi_{0} \label{scalar 2}%
\end{equation}
where $l_{\nu}=\left(  \nu^{2}-1\right)  ^{-\frac{1}{2}}$ plays the role of a
characteristic length scale of the dilaton. There is a $\pm$ ambiguity in the
integration of (\ref{scalar}), which corresponds to a discrete degeneration in
the black hole family. Indeed, from (\ref{CF}) it follows that the conformal
factor has a pole of order two at $x=1$ where the conformal infinity is
`located'. The fact that (\ref{CF}) is regular in the region $x\in\left(
0,1\right)  $ and $x\in\left(  1,\infty\right)  $ allows to pick any of these
intervals as the domain of coordinates, one corresponding to a negative scalar
and the other corresponding to a positive one.

The remaining metric function satisfies the following differential equation
(we also have $E^{\theta}_{\theta}=E^{\phi}_{\phi}$):
\begin{equation}
E_{t}^{t}-E_{\theta}^{\theta}=0\,\,\,\,\rightarrow\frac{d}{dx}\left(
\Omega(x)f^{\prime}(x)\right)  +2\eta^{2}\Omega(x)-e^{-\gamma\phi}%
Q^{2}-e^{\gamma\phi}\eta^{2}P^{2}=0 \label{FE1}%
\end{equation}
which can be exactly integrated to yield
\begin{align}
f(x)  &  =f_{0}+\frac{\eta^{2}}{\nu}\left(  x^{2}+\frac{2x^{2-\nu}}{\nu
-2}\right)  +f_{1}\left(  \frac{x^{\nu+2}}{\nu+2}-x^{2}+\frac{x^{2-\nu}}%
{2-\nu}\right) \\
&  +\frac{Q^{2}\eta^{2}}{(1-p)\nu^{2}}\left(  \frac{x^{3-p+\nu}}{3-p+\nu
}+\frac{x^{3-p-\nu}}{3-p-\nu}-2\frac{x^{3-p}}{3-p}\right)  +\frac{P^{2}%
\eta^{4}}{(1+p)\nu^{2}}\left(  \frac{x^{3+p+\nu}}{3+p+\nu}+\frac{x^{3+p-\nu}%
}{3+p-\nu}-2\frac{x^{3+p}}{3+p}\right) \nonumber
\end{align}
where $f_{0}$ and $f_{1}$ are constants of integration and $p=\gamma/l_{\nu
}=\gamma\sqrt{\nu^{2}-1}$.

From now on, we shall consider only asymptotically flat solutions. To enforce
this condition, we require
\begin{equation}
\lim_{x\rightarrow1}\Omega(x)f(x)=1 \label{BC1}%
\end{equation}
and, consequently, we fix $f_{0}$ to obtain
\begin{align}
f(x)  &  =\frac{\eta^{2}}{\nu}\left(  x^{2}+\frac{2x^{2-\nu}}{\nu-2}-\frac
{\nu}{\nu-2}\right)  +f_{1}\left(  \frac{x^{\nu+2}}{\nu+2}-x^{2}%
+\frac{x^{2-\nu}}{2-\nu}+\frac{\nu^{2}}{\nu^{2}-4}\right) \nonumber\\
&  {+}\frac{Q^{2}\eta^{2}}{(1-p)\nu^{2}}\left(  \frac{x^{3-p+\nu}}{3-p+\nu
}+\frac{x^{3-p-\nu}}{3-p-\nu}-2\frac{x^{3-p}}{3-p}-\frac{2\nu^{2}}{\left(
3-p\right)  \left(  3-p+\nu\right)  \left(  3-p-\nu\right)  }\right)
\nonumber\\
&  {+}\frac{P^{2}\eta^{4}}{(1+p)\nu^{2}}\left(  \frac{x^{3+p+\nu}}{3+p+\nu
}+\frac{x^{3+p-\nu}}{3+p-\nu}-2\frac{x^{3+p}}{3+p}-\frac{2\nu^{2}}{\left(
3+p\right)  \left(  3+p+\nu\right)  \left(  3+p-\nu\right)  }\right)
\label{general}%
\end{align}
Let us elaborate a bit more on the boundary condition (\ref{BC1}). In
canonical areal coordinates, the transverse section of the black hole should
be multiplied by $r^{2}$ at infinity. It follows from (\ref{Ansatz})
that there is a change of coordinates such that  
\begin{equation}
\Omega=r^{2}+... \label{asymptotic}%
\end{equation}
where the dots stand for subleading terms (a more precise definition
is given below). This change of coordinates is compatible with the asymptotic
form of Minkowski spacetime if the lapse function goes as 
\begin{equation}
\Omega f=1+...
\end{equation}
 which uniquely fixes (\ref{BC1}).

A detailed analysis of (\ref{general}) shows that it has regular limits when
$\nu=\pm2$, $p=\pm3$, $p=\nu\pm3$, $p=-\nu\pm3$ and it is singular at $p=\pm
1$. The singularities can be cured by picking
\begin{equation}
f_{1}=\frac{\eta^{2}Q^{2}}{\left(  p-1\right)  \nu^{2}}-\frac{\eta^{4}P^{2}%
}{\left(  p+1\right)  \nu^{2}}+\alpha+\frac{\eta^{2}(\nu+2)}{\nu^{2}}%
\end{equation}
it is easy to see that the function $f(x)$ has smooth limits for all the
values of the parameters.

This general solution is complicated, but we are going present the results for
some concrete examples corresponding to specific values of $\gamma$ only. In
particular, we are going to consider the $\gamma=\sqrt{3}$ case for which
there indeed exists a smooth limit despite the divergence in the denominator
of (\ref{general}). In the extremal case, at the horizon the metric function
should satisfy $f^{\prime\prime}(x_{h})$=constant and $f^{\prime}(x_{h})=0$.

Once we obtain the metric, it is straightforward to algebraically
find the potential so that the field equations are satisfied. In
general, the potential is a function of six parameters: besides
$\gamma$, $\nu$, $Q$, and $P$ it is possible to introduce a
cosmological constant and an extra parameter, $\alpha$, which
appears as an overall multiplication constant in the potential,
after a redefinition of $f_{1}$.

A general class of solutions with the gauge fields turned on and with
cosmological constant is going to be presented in \cite{noi}. In what follows,
we focus on asymptotically flat solutions when the gauge fields are turned on
and there is a non-trivial dilaton potential --- for these solutions the
dilaton potential is vanishing at the boundary. With the ansatz (\ref{Ansatz}%
), it is possible to construct a five parameter family of dyonic
black hole solutions.

Dyonic black holes for the Einstein-Maxwell-Dilaton theory that can be
embedded in string theory are known for $\gamma=1$ and $\gamma=\sqrt{3}$. In
what follows, we present the extension of the $\gamma=1$ solution. For
$\gamma=\sqrt{3}$, we have also found a dyonic black hole. However, in this
case, the potential has a part that is proportional with the magnetic charge
and the only way to consider the limit of vanishing potential is when the
magnetic charge is also vanishing. In this limit, the solution matches the
\textit{electrically} charged Kaluza-Klein black hole. We are going to also
present a generalization of `hairy' Reissner-Nordstr{\"{o}}m black holes for
which the scalar field and gauge field are not coupled, but the dilaton
potential is non-trivial.

More general solutions are listed in Appendices \ref{SUGRA1} and \ref{SUGRA2}
but, for simplicity, in this section we are going to focus only on some
deformations of solutions that can be embedded in supergravity.

\subsection{Solutions with a non-trivial dilaton potential}

\subsubsection{$\gamma=1$}

To obtain this solution we should also consider the limit $\nu\rightarrow
\infty$ --- a more general family of solutions with arbitrary $\gamma$ is
presented in Appendix \ref{SUGRA1}.

To obtain the limit $\nu\rightarrow\infty$ in a smooth way, we change the
coordinates as \thinspace%
\begin{equation}
x\rightarrow x^{\frac{1}{\nu}},\qquad t\rightarrow\frac{t}{\nu},
\end{equation}
and rescale the $2$-sphere as
\begin{equation}
d\theta^{2}+\sin^{2}\theta d\varphi^{2}\rightarrow\nu^{-2}\left(  d\theta
^{2}+\sin^{2}\theta d\varphi^{2}\right)  , \label{rescaling}%
\end{equation}
in which case the $\nu^{2}$ of the conformal factor (\ref{CF}) cancels out.

After taking the limit, we obtain
\begin{equation}
\Omega(x)=\frac{x}{\eta^{2}\left(  x-1\right)  ^{2}} \,\,\,\,\,\,\,\,\, ,
\,\,\,\,\,\, \qquad\phi(x)=\ln(x) \label{dilaton}%
\end{equation}%
\begin{equation}
ds^{2}=\Omega(x)\left[  -f(x)dt^{2}+\frac{\eta^{2}dx^{2}}{x^{2}f(x)}%
+d\theta^{2}+\sin^{2}{\theta}d\varphi^{2}\right]
\end{equation}

We emphasize that the $g_{xx}$ metric component is now different
than the one for finite $\nu$ (see (\ref{Ansatz})). One can work
with an arbitrary parameter $\gamma$ in the dilatonic coupling. The
field equations solvable and the solution is presented in Appendix
A. However, in this section, we describe only the $\gamma=1$ case,
which is very simple and can be smoothly connected with a solution
of $\mathcal{N}=4$ supergravity. To completely characterize the
solution, apart from the conformal factor and dilaton
(\ref{dilaton}), we have the following remaining data:
\begin{equation}
A=\frac{Q}{x}dt+P\cos\theta d\varphi\label{A}%
\end{equation}%
\begin{equation}
V(\phi)=\alpha\left[  2\phi+\phi\cosh(\phi)-3\sinh(\phi)\right]
\label{potential1}%
\end{equation}%
\begin{equation}
f(x)=\frac{\eta^{2}(x-1)^{2}}{x}+\left[  \frac{x}{4}-\frac{1}{4x}-\frac{1}%
{2}\ln(x)\right]  \alpha+\frac{\eta^{2}(x-1)^{3}}{2x}\left(  \eta^{2}%
P^{2}-x^{-1}Q^{2}\right)
\end{equation}
One can easily see that at the boundary, $x=1$, the dilaton is vanishing and
so does the dilaton potential. The limit for vanishing dilaton potential is
obtained when $\alpha=0$.

Let us now show explicitly that the metric is asymptotically flat by studying
the boundary $x=1$. To do this we look for a change of coordinates such that
at infinity the sphere is multiplied by $r^{2}$
\begin{equation}
\Omega(x)=r^{2}+O(r^{-4}), \label{AF}%
\end{equation}
which is given for, the $x<1$ black holes, by
\begin{equation}
x=1-\frac{1}{\eta r}+\frac{1}{2\eta^{2}r^{2}}-\frac{1}{8\eta^{3}r^{3}}%
+\frac{1}{2^{7}\eta^{5}r^{5}} \label{CC}%
\end{equation}

Now, it is possible to see that the metric is asymptotically flat%

\begin{equation}
g_{tt}=\Omega(x)f(x)=1-\frac{\alpha+6\eta^{2}\left(  \eta^{2}P^{2}%
-Q^{2}\right)  }{12\eta^{3}r}+O(r^{-2})
\end{equation}

\begin{equation}
g_{rr}^{-1}=\frac{x^{2}f(x)}{\eta^{2}\Omega(x)}\left(  \frac{dr}{dx}\right)
^{2}=1-\frac{\alpha+6\eta^{2}\left(  \eta^{2}P^{2}-Q^{2}\right)  }{12\eta
^{3}r}+O(r^{-2}).
\end{equation}

The same change of coordinates exists for $x>1$. The scalar field potential is
everywhere regular, except at the spacetime singularities.

\subsubsection{$\gamma=\sqrt{3}$}

Let us consider now another interesting solution, which can be smoothly
connected with the electrically charged Kaluza-Klein black hole when the
dilaton potential is vanishing. This case corresponds to
\begin{equation}
\gamma=\left(  \frac{\nu+1}{\nu-1}\right)  ^{1/2}%
\end{equation}
when $\nu=2$ --- once again, we present a more general family of solutions
with $\gamma$ arbitrary in Appendix \ref{SUGRA2}. The metric is
\begin{equation}
ds^{2}=\Omega(x)\left[  -f(x)dt^{2}+\frac{\eta^{2}dx^{2}}{f(x)}+d\theta
^{2}+\sin^{2}\theta d\varphi^{2}\right]
\end{equation}
where the metric function, $f(x)$, and conformal factor, $\Omega(x)$, are
\begin{equation}
f(x)=\alpha\left[  2\ln(x)+\frac{x^{4}}{2}+\frac{3}{2}-2x^{2}\right]
+\frac{\eta^{4}P^{2}}{384}\left(  x^{2}-1\right)  ^{3}\left(  3x^{2}+1\right)
-\frac{\eta^{2}(x^{2}-1)^{3}Q^{2}}{16x^{2}}+\frac{\eta^{2}(x^{2}-1)^{2}}{4}%
\end{equation}
\begin{equation}
\Omega(x)=\frac{4x}{\eta^{2}\left(  x^{2}-1\right)  ^{2}}%
\end{equation}
The dilaton and gauge potential are
\begin{equation}
\phi=\sqrt{3}\ln(x) \,\,\,\,\,\,\,\,\, , \,\,\,\,\,\,\,\, A=-\frac{Q}{2x^{2}%
}dt+P\cos{\theta}d\varphi
\end{equation}
and the dilaton potential is
\begin{equation}
V(\phi)=\alpha\left[  \sinh(\sqrt{3}\phi)+9\sinh(\frac{\phi}{\sqrt{3}}%
)-\frac{12\phi}{\sqrt{3}}\cosh(\frac{\phi}{\sqrt{3}})\right]  -\frac
{e^{-\frac{\phi}{\sqrt{3}}}\eta^{4}P^{2}\left(  e^{\frac{2\phi}{\sqrt{3}}%
}-1\right)  ^{5}}{2^{7}}%
\end{equation}
Again we observe that at the boundary $x=1$ the dilaton is vanishing, and the
dilaton potential likewise vanishes there. However, the limit for which the
dilaton potential is vanishing corresponds this time to $\alpha=0$ and $P=0$
--- in this limit we recover the electrically charged Kaluza-Klein black hole.
One important observation is that, unlike the extremal limit of an
electrically charged Kaluza-Klein black hole that is singular, our solution
has a regular extremal limit due to the presence of the dilaton potential. We
are going to discuss this point in more detail in Section $4$.

\subsubsection{Hairy Reissner-Nordstr{\"o}m black hole}

We consider the theory for which gauge coupling does not depend of the
dilaton, namely $\gamma=0$. As follows from the equations of motion, this is a
consistent truncation of the non-minimally coupled theory. Therefore,
following the results of Section $2$, the scalar field and the conformal
factor are given by (\ref{scalar 2}) and (\ref{CF}), respectively. The
integration of the field equation for $f(x)$, is given by the expression
(\ref{general}) when $\gamma=p=0$ and yields%
\begin{align}
f(x)  &  =\frac{\eta^{2}}{\nu}\left(  x^{2}+\frac{2x^{2-\nu}}{\nu-2}-\frac
{\nu}{\nu-2}\right)  +f_{1}\left(  \frac{x^{\nu+2}}{\nu+2}-x^{2}%
+\frac{x^{2-\nu}}{2-\nu}+\frac{\nu^{2}}{\nu^{2}-4}\right) \nonumber\\
&  {+}\frac{\left(  Q^{2}\eta^{2}+P^{2}\eta^{4}\right)  }{\nu^{2}}\left(
\frac{x^{3+\nu}}{3+\nu}+\frac{x^{3-\nu}}{3-\nu}-2\frac{x^{3}}{3}-\frac
{2\nu^{2}}{3\left(  3+\nu\right)  \left(  3-\nu\right)  }\right)
\end{align}
which upon using the definitions $q=\frac{2\eta^{2}\left(  Q^{2}+\eta^{2}%
P^{2}\right)  }{3}$, $f_{1}=\frac{\alpha}{\nu^{2}}+\frac{\eta^{2}\left(
\nu+2\right)  }{\nu^{2}}$ yields%
\begin{align}
f(x)  &  =\frac{x^{2-\nu}\eta^{2}}{\nu^{2}}(x^{\nu}-1)^{2}+\alpha\left[
\frac{1}{\nu^{2}-4}-\left(  1+\frac{x^{-\nu}}{\nu-2}-\frac{x^{\nu}}{\nu
+2}\right)  \frac{x^{2}}{\nu^{2}}\right] \nonumber\\
&  +q\left[  \frac{1}{\left(  \nu^{2}-9\right)  }-\left(  1+\frac{3}{2}%
\frac{x^{-\nu}}{\nu-3}-\frac{3}{2}\frac{x^{\nu}}{3+\nu}\right)  \frac{x^{3}%
}{\nu^{2}}\right]
\end{align}

Unlike the previous case, the dilaton potential has a part that depends on
both electric and magnetic charges, through the combination $q=\frac{2\eta
^{2}\left(  Q^{2}+\eta^{2}P^{2}\right)  }{3}$, which appears also in the
metric. The dilaton potential is more complicated:
\begin{align}
V(\phi)  &  =\frac{2\alpha}{\nu^{2}}\left(  \frac{\nu-1}{\nu+2}\sinh(\left(
1+\nu\right)  \phi l_{\nu})+\frac{\nu+1}{\nu-2}\sinh(\left(  1-\nu\right)
\phi l_{\nu})+4\frac{\nu^{2}-1}{\nu^{2}-4}\sinh\left(  \phi l_{\nu}\right)
\right) \nonumber\\
&  +\frac{3qe^{2\phi l_{\nu}}}{4\nu^{4}}\left(  \frac{\nu-1}{\nu+3}e^{2\nu\phi
l_{\nu}}+\frac{\nu+1}{\nu-3}e^{-2\nu\phi l_{\nu}}\right) \nonumber\\
&  +\frac{q\left(  \nu^{2}-1\right)  e^{2\phi l_{\nu}}}{\nu^{4}}\left(
\frac{\nu-3}{\nu+3}e^{\nu\phi l_{\nu}}+\frac{\nu+3}{\nu-3}e^{-\nu\phi l_{\nu}%
}+\frac{1}{2}\frac{\left(  8\nu^{2}-27\right)  }{\left(  \nu^{2}-9\right)
}\right) \nonumber\\
&  -\frac{q(\nu^{2}-4)e^{-\phi l_{\nu}}}{\nu^{2}\left(  \nu^{2}-9\right)
}\left(  \frac{\nu-1}{\nu+2}e^{-\nu\phi l_{\nu}}+\frac{\nu+1}{\nu-2}e^{\nu\phi
l_{\nu}}-4\frac{\left(  \nu^{2}-1\right)  }{(\nu^{2}-4)}\right)  ,
\end{align}
and it has been arranged such that it is explicitly invariant under
$\nu\rightarrow-\nu$. The gauge field is given by (\ref{GF}) with $\gamma=0$.

Let us close this section with a couple of comments. We observe, as expected,
that a consistent limit for which the Reissner-Nordstr{\"o}m solution is
recovered is $\nu=1$ (which also means $\phi=0$) when the dilaton potential is
vanishing. The potential vanishes when $\alpha=0$ and both charges also vanish
(because we should have $q$=0 so that the part of the potential proportional
to $q$ is also vanishing). In other words, there are no regular hairy
Reissner-Nordstr{\"o}m black holes with a vanishing dilaton potential, as
expected from no-hair theorems.

\subsection{Solutions with a vanishing dilaton potential}

The solutions we found are regular and generalise some of the
solutions that were previously known for the case of vanishing
scalar potential. The solutions presented in \cite{Gibbons:1987ps,
Dobiasch:1981vh, Gibbons:1985ac, Astefanesei:2006sy} were obtained
by rewriting the equations of motion as Toda equations
\cite{Gibbons:1987ps}. This method is useful for obtaining solutions
in certain cases with exponential couplings. However, it is not
powerful enough to construct more general solutions with non-trivial
moduli potentials.

We are interested in enough simple solutions (with regular supersymmetric
limits), but a whole new family of solutions can be generated from our general
solutions presented in Appendices.

We are going to explicitly show how an old known solution can be recovered
from the new solutions with a non-trivial potential that we have constructed.
This solution was presented in \cite{Kallosh:1992ii} (see, also,
\cite{Astefanesei:2006sy} for a detailed discussion of its thermodynamical
properties and the extremal limit in the context of attractor mechanism) and
has the following form:%
\begin{equation}
ds^{2}=-a(r)^{2}dt^{2}+a(r)^{-2}dr^{2}+b(r)^{2}d\Omega^{2} \label{metric2}%
\end{equation}%
\begin{equation}
\exp(2\phi)=\frac{r+\Sigma}{r-\Sigma}\,\,,\,\,a^{2}=\frac{(r-r_{+})(r-r_{-}%
)}{r^{2}-\Sigma^{2}}\,\,,\,\,b^{2}=r^{2}-\Sigma^{2} \label{sol1}%
\end{equation}
with
\begin{equation}
r_{\pm}=M\pm\sqrt{M^{2}+\Sigma^{2}-Q^{2}-P^{2}}%
\end{equation}
We emphasize that the scalar charge $\Sigma$ is not an independent parameter:
\begin{equation}
\Sigma=\frac{P^{2}-Q^{2}}{2M}%
\end{equation}
The scalar charge plays a role in the first law when the asymptotic value of
the dilaton does vary \cite{Gibbons:1996af}. The extremal limit of the above
solution is obtained by letting $r_{+}=r_{-}$ and the corresponding solution
can be embedded in $\mathcal{N}=4$ supergravity.

We can reobtain this solution from our solution $\gamma=1$ when the dilaton
potential is vanishing ($\alpha=0$). To find the right change of coordinates,
first we match the scalars and we get:
\begin{equation}
x=\frac{r+\Sigma}{r-\Sigma}\Longrightarrow\phi=\ln(\frac{r+\Sigma}{r-\Sigma
})\text{,}%
\end{equation}

Then, to obtain the same expressions for the gauge field, we redefine
$Q=-\frac{q}{2\Sigma}$ in (\ref{A}) and so we get the following gauge
potential:
\begin{equation}
A=-\frac{q}{2\Sigma}\frac{r-\Sigma}{r+\Sigma}dt+P\cos\theta d\varphi
\end{equation}
The last step is to fix $\eta$ by matching the $b^{2}$ of (\ref{metric2}) with
the conformal factor $\Omega$:
\begin{equation}
\Omega(x(r))=r^{2}-\Sigma^{2},\qquad\eta=\frac{1}{2\Sigma}%
\end{equation}
Once we have the right change of coordinates
\begin{equation}
dx=-\frac{2\Sigma dr}{(r-\Sigma)^{2}},\qquad\frac{\eta^{2}dx^{2}}{x^{2}}%
=\frac{dr^{2}}{(r^{2}-\Sigma^{2})^{2}}%
\end{equation}
it is straightforward to obtain
\begin{equation}
a(r)=\frac{4\Sigma r^{2}+(P^{2}-q^{2})r+\Sigma q^{2}+\Sigma P^{2}-4\Sigma^{3}%
}{4\Sigma(r^{2}-\Sigma^{2})}%
\end{equation}
which matches (\ref{sol1}) when the charges are rescaled with a factor of
$1/2$ as it should be for a $2$-form field strength.

One can similarly show that the $\gamma=\sqrt{3}$ is smoothly connected with
the electrically charged Kaluza-Klein black hole (that can be obtained by
Kaluza-Klein reduction of the $5d$ Schwarzschild black hole) of
\cite{Dobiasch:1981vh,Gibbons:1985ac}.

\bigskip

\section{Hair and thermodynamical properties.}

\label{hairyy}

To study the first law of thermodynamics for our black holes, we first compute
the entropy and the temperature. In the coordinates we use to write our metric
ansatz (\ref{Ansatz}), the boundary is at $x=1$ and the horizon is at
$x=x_{h}$. The temperature (an intensive quantity) is conformally invariant
and so depends only on $f(x)$ and not on the conformal factor:
\begin{equation}
T=\frac{f^{\prime}(x)}{4\pi\eta}\bigg|_{x=x_{h}}%
\end{equation}
The entropy and mass can be also easily computed and we get
\begin{equation}
S=\frac{\pi\Omega(x_{h})}{G_{N}}\,\,\, , \,\,\,\,M=\frac{\left[  \Omega(x)
f(x)\right]  ^{\prime}}{2\eta G_{N}}\bigg|_{x=1}%
\end{equation}
To gain some intuition, let us first discuss the Schwarzschild solution
($\nu=-1$ in Appendix B) in the coordinates we use
\begin{equation}
\Omega=\frac{1}{\eta^{2} (x-1)^{2}} \,\,\,\,\, , \,\,\,\,\,\, f(x)=(x-1)^{2}
\left[  \eta^{2}+ (x-1)\left(  \eta^{2} + \frac{\alpha}{12}\right)  \right]
\end{equation}
for which $x_{h} = \frac{\alpha}{\alpha+ 12\eta^{2}}$. The change of
coordinates to put the metric in the canonical Schwarzschild form is
\begin{equation}
x=1-\frac{1}{\eta r}%
\end{equation}
Using the general formulas for the temperature, entropy, and mass presented
above, we obtain
\begin{equation}
M=\frac{1}{2\eta G_{N}} \left(  1+\frac{\alpha}{12\eta^{2}} \right)
\,\,\,\,\, , \,\,\,\,\, S=\frac{\pi}{G_{N}}\frac{1}{\eta^{2} (x_{h}-1)^{2}%
}=\frac{\pi}{G_{N}}\frac{1}{\eta^{2}}\left(  1+\frac{\alpha}{12\eta^{2}}
\right)  ^{2}%
\end{equation}
\begin{equation}
T=\frac{ (x_{h}-1)\left[  8\eta^{2}+(x_{h}-1)( \alpha+12\eta^{2}) \right]
}{16\pi\eta} = \frac{ \eta}{4\pi} \left(  1+\frac{\alpha}{12\eta^{2}} \right)
^{-1}%
\end{equation}
from which we recover the well known results
\begin{equation}
M= \frac{m}{G_{N}} \rightarrow S=\frac{4\pi m^{2}}{G_{N}} , T=\frac{1}{8\pi m}%
\end{equation}
in the limit $\alpha\to0$.

The only subtlety related to the first law in the coordinates we use comes
from the fact that the horizon radius is dimensionless. However, the
computations above can be straightforwardly generalized to our solutions. For
concreteness, let us now discuss the $\gamma=\sqrt{3}$ solution. Similar but
more involved computations can be done for the other solutions we have
presented in the previous section.

With our conventions, the electric charge can be written as
\begin{equation}
q=-\frac{Q}{4\eta G_{N}}%
\end{equation}
and the conjugate potential is $\Phi_{q}=\frac{Q}{2}\left(  1-\frac{1}%
{x_{h}^{2}} \right)  $ (which is, up to an additive constant, just the usual
gauge potential $\Phi_{q}=Q/x_{h}^{2}$) in a gauge such that it vanishes at
the boundary $x=1$ and it is finite at the horizon.

The expressions for the mass, temperature, and entropy for the $\gamma
=\sqrt{3}$ solution are
\begin{equation}
M=\frac{32\alpha+ 12\eta^{2} - 6\eta^{2}Q^{2}+ \eta^{4}P^{2}}{24\eta^{3}G_{N}}%
\end{equation}
\begin{equation}
T=\frac{(x_{h}^{2}-1)^{2}}{4\pi\eta} \left[  \frac{2\alpha}{x_{h}}+\frac
{\eta^{4}x_{h}^{3}P^{2}}{16} - \frac{\eta^{2}(2x_{h}^{2}+1)Q^{2}}{8x_{h}^{3}}
+ \frac{\eta^{2} x_{h}}{x_{h}^{2}-1} \right]
\end{equation}

\begin{equation}
S=\frac{4\pi x_{h}}{\eta^{2}(x_{h}^{2}-1)^{2}G_{N}}%
\end{equation}

For simplicity, one can consider the case $P=0$. Since the radius of the
horizon, $x_{h}$, cannot be explicitly obtained from $f(x_{h})=0$, one can
verify the first law by using the chain rule to get the variations with
respect to the independent parameters $Q$ and $\eta$:
\begin{equation}
dM(q, \eta)=TdS(q,\eta) + \Phi_{q} dq
\end{equation}

\section{Extremal limit and attractor mechanism.}

\label{attractor}

The role of attractor mechanisms \cite{Ferrara:1995ih} for computing the
entropy of extremal (supersymmetric or non-supersymmetric) black holes is by
now well understood \cite{Dabholkar:2006tb, Sen:2007qy, Astefanesei:2006sy}.
Here we would like to present an interesting result on the existence of
extremal black holes: we find that there are \textit{regular} black hole
solutions in a theory with only one gauge field (electric or magnetic)
non-minimally coupled with the dilaton for non-trivial dilaton potentials. The
attractor mechanism can be used then as a criterion for the existence of
regular extremal black hole solutions. Similar considerations were made for
baryonic and electromagnetic branes and other solutions in $AdS$ spacetime in
\cite{Astefanesei:2010dk}. More importantly, since the extremal (attractor)
horizons are infinitely far away in the bulk\footnote{It was shown in
\cite{Kunduri:2007vf} that the near horizon geometry of extremal stationary
black holes contains an $AdS_{2}$ spacetime.} they do not get distorted by
changing the boundary values of the scalars and so, from this point of view,
the attractor mechanism acts as a no-hair theorem \cite{Israel:1967wq} for
extremal black holes \cite{Astefanesei:2007vh}.

To understand why is so, let us review some basic facts related to the
effective potential \cite{Goldstein:2005hq}. Since the dilaton is
non-minimally coupled with the gauge field, a new term appears in its equation
of motion. This term, which is controlled by the gauge field, can be
interpreted as an \textit{effective potential} for the dilaton. For example,
for the spherically symmetric ansatz (\ref{metric2}), an electromagnetic field
with both electric and magnetic charges\footnote{For two electric fields, to
get regular solutions when the dilaton potential is turned off, the gauge
couplings should be different so that the effective potential has a minimum at
the horizon.} turned on and the exponential coupling $e^{\phi}$, the effective
potential is
\begin{equation}
V_{eff}(\phi)=\frac{1}{b^{2}}\left[  e^{-\phi}Q^{2} + e^{\phi}P^{2} \right]  +
2b^{2} V(\phi)
\end{equation}

Let us first discuss the case when the dilaton potential is vanishing
($V(\phi)=0$). We observe that when just one charge is non-zero, the effective
potential does not have a minimum at the horizon, which means that that there
are no regular solutions in this case. When a second charge is turned on, the
effective potential has a minimum and regular solutions exist. When the
dilaton potential is non-zero, since there is a competition between the part
generated by the gauge fields and pure dilaton part, the situation can
drastically change. That is, there exist regular solutions even when just one
charge is non-zero.

In principle, even if technically difficult, one could try to find some
general conditions on the moduli potential so that the theory contains regular
extremal solutions. However, such an analysis is beyond the goal of this paper
and we prefer to prove the existence of this kind of solutions using the
entropy function formalism of Sen \cite{Sen:2005wa} (see, also,
\cite{Astefanesei:2006dd} for stationary black holes). The important point is
that the effective potential method and the entropy function formalism are
equivalent in the near horizon limit.\footnote{Since we have exact solutions,
we could find the near horizon metric directly for each solution. However such
a computation is difficult and not very illuminating.} Since the near horizon
geometry of extremal black holes has an enhanced symmetry of $AdS_{2} \times
S^{2}$ (in the extremal case, the near horizon geometry is a solution of the
equations of motion), the computations simplify drastically in the near
horizon limit.

In what follows, we discuss the simplest potential (\ref{potential1}), but the
same kind of argument applies to more general potentials. The ansatz for the
near horizon geometry of $4-$dimensional static extremal black holes is
\begin{equation}
ds^{2}=v_{1}(-r^{2}dt^{2} + \frac{dr^{2}}{r^{2}}) + v_{2}\left[  \frac{dy^{2}%
}{1-y^{2}} + (1-y^{2})d\phi^{2} \right]
\end{equation}
By solving the attractor equations one can obtain the near horizon data,
$\{v_{1}, v_{2}\}$, and also the horizon value of the dilaton ($u$) as
functions of the physical charge $q$ and the parameter $\alpha$ in the
potential. After some straightforward computations we find
\begin{equation}
\label{eqalpha}\alpha=8\,\frac{e^{u}}{q^{2}}\,\frac{2+u\sinh{u}-2\cosh{u}
}{[2(1+u)+u(\sinh{u}+\cosh{u})-2\cosh{u}-3\sinh{u}]^{2}}%
\end{equation}
We cannot solve this equation explicitly to obtain the horizon value of the
dilaton. However for our purposes it is sufficient to observe that since the
parameter $\alpha$ is arbitrary, we have the freedom to fix it so that the
corresponding horizon value of the dilaton and the radii of $AdS_{2}$ and
sphere are well defined at the horizon and so the solution is regular.

\section{Discussion}

\label{discuss} In this paper we obtained asymptotically flat static solutions
to the Einstein-Maxwell-dilaton system in the presence of a non-vanishing
dilaton potential. Solving the field equations subject to a special ansatz for
the metric and dilaton fields, we found regular solutions. Some of these
contain examples that were considered previously in the literature and we have
verified that our solutions (smoothly) reduce to these in the appropriate
corners of parameter space. At this point, it is not clear to us how (or if it
is possible) to embed our solutions into supergravity; this is an interesting
open question that we leave for the future.

The no-hair theorem implies that short-range interactions decaying fast enough
at infinity (hair) are not allowed in the exterior of a stationary black hole.
Since the hair does not contribute to Gauss-like conserved charges, the
classical degrees of freedom of the black hole are restricted to those related
to its conserved charges. Over the past decade the accuracy of this theorem
was strongly challenged. Indeed, hairy black hole solutions were discovered
\cite{Torii:2001pg, Sudarsky:2002mk, Anabalon:2012ta} that appear to evade the
no-hair theorems. Interestingly enough, we were able to obtain \textit{exact}
hairy asymptotically flat black hole solutions similar to the one predicted in
\cite{Sudarsky:2002mk}. To get a better understanding, let us now compare one
of our dilaton potentials, the simplest one presented in (\ref{potential1}),
with a generic potential described in \cite{Sudarsky:2002mk}. The main
assumption in \cite{Sudarsky:2002mk} is that the family of dilaton potentials
used to construct the solution are not positive-semidefinite and the weak
energy condition is not satisfied. Also, to obtain asymptotically flat
solutions, the dilaton settles asymptotically at the local \textit{minimum},
which should also correspond to a root of the potential. Therefore, the
conditions imposed on the potentials of \cite{Sudarsky:2002mk} for the
existence of the asymptotically flat hairy black hole solutions are such that
a root and a local minimum of the potential are located at the same place:
$V(\phi_{\infty})=\partial_{\phi} V(\phi)|_{\phi_{\infty}}=0$ and
$\partial^{2}_{\phi\phi} V(\phi)|_{\phi_{\infty}}>0$. Now, let us consider the
dilaton potential (\ref{potential1}) and study its asymptotic behaviour for
$\phi_{\infty}=0$. We obtain: $V(0)=\partial_{\phi} V(\phi)|_{\phi=0}%
=\partial^{2}_{\phi\phi} V(\phi)|_{\phi=0}=0$ and so the potential becomes
very flat along this direction (in fact, the third and fourth derivatives
vanish also at $\phi=0$). However, although the second derivative of the
potential vanishes, the fifth does not and so the dilaton only feels the fifth
derivative of the potential. We would like to point out that this potential
resembles the potential of the Minimally Supersymmetric Standard Model (MSSM)
flat direction inflation models (see, e.g., \cite{Allahverdi:2006we}%
).\footnote{The fields that compose flat directions may thus in principle be
excited to large classical field strengths at no cost to the potential energy.
Within MSSM, for example, all the flat directions are lifted by
non-renormalizable operators.}

When a classical scalar field acts as a source of gravity, the energy
conditions can be violated depending on the form of the scalar potential and
the value of the curvature coupling \cite{Barcelo:2000zf}. Let us consider
again the simplest case (\ref{potential1}). The sign of the potential is
controlled by the sign of the parameter $\alpha$ --- this can be easily seen
by considering the derivatives in the Taylor expansion (close to the boundary)
of the part of the potential that multiplies $\alpha$ and observing that all
these derivatives are positive or zero. Therefore, a positive potential
corresponds to a positive $\alpha$. However, we can explicitly check that, in
this case and for $Q=P=0$, there are no black hole solutions, there exist just
naked singularities. When $\alpha$ is negative and $Q=P=0$ there exist regular
black hole solutions but the weak energy condition is violated in agreement
with the results of \cite{Barcelo:2000zf}. On the other hand, when $(Q,P)$ are
non-zero, there exist regular black hole solutions even for a positive
$\alpha$ and the weak energy condition is satisfied in this case.

The null energy condition is always satisfied. The energy momentum of the
scalar field, in a comoving tetrad, has the form $T^{ab}=diag(\rho,p_{1}%
,p_{2},p_{2})$ and satisfies the null energy condition. For our static
solutions with the condition $f(x)>0$, we obtain
\begin{equation}
\rho+p_{2}=0,\qquad\rho+p_{1}=\frac{\left(  \nu^{2}-1\right)  \left(  x^{\nu
}-1\right)  ^{2}f(x)}{2\nu^{2}x^{\nu+1}}>0.
\end{equation}

For the non-extremal black holes, the near horizon geometry and the
entropy vary continuously as the asymptotic values of moduli fields
are changing. For example, we saw that for our solutions, the
horizon radius depends of the horizon value of the dilaton which is
controlled by the asymptotic data. On the other hand, for the
extremal black holes, the moduli fields in a black hole background
vary radially and get attracted to certain specific values at the
horizon, which in fact depend only on the quantized charges of the
black hole under consideration. It is then tempting to interpret
that the attractor mechanism plays the role of a no-hair theorem for
the extremal black holes \cite{Astefanesei:2007vh}. As we have
already described in detail in Section \ref{attractor}, due to the
competition between the effective potential and dilaton potential,
there exist also extremal black hole solutions with only one gauge
field turned on.

We would like to comment now on the meaning of changing the dilaton's
asymptotic value when the dilaton potential does not vanish. The equation
(\ref{FE1}) will have the same expression if we use the `scaled' charges
$\bar{Q}=e^{-\gamma\phi_{\infty}}$ and $\bar{P}=e^{\gamma\phi_{\infty}}$ and
so the solutions have the same form if we replace the conserved charges with
the `scaled' charges. First, let us review what is happening when the dilaton
potential vanishes \cite{Gibbons:1996af}. Due to the non-minimal coupling of
the dilaton, the first law should be supplemented by a new term containing the
variation of the asymptotic value of the dilaton: $\Sigma\, d\phi_{\infty}$.
Here, $\Sigma$ is the scalar charge that is defined as the monopole in the
multipole expansion of the dilaton at infinity and $\phi_{\infty}$ is the
asymptotic value of the dilaton. However, the correct interpretation for the
existence of such a term was presented in \cite{Astefanesei:2006sy}. That is,
the new term does not have a similar interpretation like the other terms in
the first law because it corresponds to a change in the couplings (so, by the
variation of $\phi_{\infty}$ a different theory is obtained rather than
another thermodynamic configuration). If one tries to use the scaling symmetry
to get rid of this term, a different problem is encountered: the scaling
symmetry does not preserve the conserved charges. The new solution cannot be
reached \textit{dynamically} starting from the old one because this will also
force a violation of charge conservation. When the dilaton potential is not
vanishing there is an even more drastic change. That is, by varying the
asymptotic value of the dilaton, its potential is changing and, consequently,
that effect corresponds to a change in the asymptotics of the solution.

Even if supersymmetry is experimentally discovered, one still has to
understand how is it broken in the nature. It is possible to
construct models for which the supersymmetry is broken spontaneously
or by hand (as in our case). We have obtained solutions that are
\textit{smoothly} connected to some sugra solutions for which the
microscopic entropy and the holographic (CFT) degrees of freedom are
well understood. A related speculative observation is that it may be
possible that our solutions could be described holographically by a
dual QFT, which in some limit becomes a CFT and this is very similar
to an RG flow towards a fixed point.

One important question is if these black holes are stable. Such an analysis is
beyond the scope of this work. However, we would like to point out that in the
extremal case the black holes are long lived states (metastable) that
correspond to the minimum mass of the theory they belong to --- the
non-perturbative effects should play, in fact, an important role for
understanding the stability of extremal black holes. This argument together
with the attractor mechanism are at the basis of computing the microscopic
entropy of extremal non-supersymmetric black holes in string theory
\cite{Dabholkar:2006tb, Astefanesei:2006sy}.

For future directions, one can study the phase structure of these black holes.
This can be done, for example, by computing the quasilocal stress-tesor,
action, and relevant thermodynamical quantities as in
\cite{Astefanesei:2005ad} Among other possible applications, one could find
and describe the so-called `hairosphere' \cite{Nunez:1996xv}. We have
explicitly shown that the non-linear character of the matter fields plays a
key role in the construction of hairy black hole configurations. One
interesting question is how close to a black hole event horizon can matter
hover such that the non-linear behavior of the hair is present (in other
words, how `short' can hair be) \cite{Nunez:1996xv}?\footnote{In the
asymptotic region, the behavior of the fields is dominated by linear terms in
their respective equations of motion.} Since we have exact solutions, another
possible application is to check the `no-short-hair' conjecture of
\cite{Nunez:1996xv}.

We would like to emphasize that with our method we can also generate
solutions with scalars (and gauge fields) when the cosmological
constant is positive --- these solutions can be described along the
lines of \cite{Leblond:2002ns} (e.g., computing the corresponding
c-functions). These solutions could provide interesting starting
points for building novel cosmological scenarios.

\section{Acknowledgments.}

We would like to thank Andres Acena for important discussions and
collaboration on related projects. DA would also like to thank David Choque
for interesting conversations. Research of A.A. is supported in part by the
Fondecyt Grant 11121187. The work of DA is supported by the Fondecyt Grant
1120446. This work was supported in part by the Natural Sciences and
Engineering Research Council of Canada.

\appendix{}


\section{A general family of dilatonic dyonic black holes}

\label{SUGRA1}
Let us consider the following action
\begin{equation}
I[g_{\mu\nu},A_{\mu},\phi]=\frac{1}{2\kappa}\int d^{4}x\sqrt{-g}\left[
R-\frac{1}{4}e^{\lambda\phi}F^{2}-\frac{1}{2}\partial_{\mu}\phi\partial^{\mu
}\phi-V(\phi)\right]  \label{actionA}%
\end{equation}

The general class of solutions we present here is for $\lambda$ arbitrary and
$\nu\rightarrow\infty$ and contains the solution presented in Section $2.1.1$,
which corresponds to $\lambda=1$. To obtain regular solutions in the limit
$\nu\rightarrow\infty$ we should again rescale the coordinates as in Section
$2.1.1$ to obtain the following metric ansatz:
\begin{equation}
ds^{2}=\Omega(x)\left[  -f(x)dt^{2}+\frac{\eta^{2}dx^{2}}{x^{2}f(x)}%
+d\theta^{2}+\sin^{2}{\theta}d\varphi^{2}\right]
\end{equation}
where again the conformal factor is
\begin{equation}
\Omega(x)=\frac{x}{\eta^{2}(x-1)^{2}}%
\end{equation}
With this form of the conformal factor, it is easy to obtain the dilaton and
the $U(1)$ field strength:
\begin{equation}
\phi=\ln(x)\,\,\,\,\,\,\,\,,\,\,\,\,\,\,\,\,A=\frac{Q}{x^{\lambda}%
}dt+Pyd\varphi
\end{equation}
However, the dilaton potential and the metric are more complicated in this
case. Let us first define the following expressions:
\begin{equation}
\alpha_{1}=\eta^{2}(P^{2}\eta^{2}-\lambda^{2}Q^{2}),\qquad\alpha_{2}=\eta
^{2}(P^{2}\eta^{2}+\lambda^{2}Q^{2}),
\end{equation}

\begin{align}
W(h(\phi))  &  =-\frac{\left(  3\lambda^{4}-9\lambda^{2}+8\right)
h(\lambda\phi)}{\left(  \lambda^{2}-1\right)  \lambda^{2}}-\frac{\left(
\lambda+1\right)  h(\left(  \lambda-2\right)  \phi)}{2\left(  \lambda
-1\right)  }-\frac{\left(  \lambda-1\right)  h(\left(  \lambda+2\right)
\phi)}{2\left(  \lambda+1\right)  }\nonumber\\
&  +\frac{2\left(  \lambda^{2}+\lambda-1\right)  \left(  \lambda-1\right)
h(\left(  \lambda+1\right)  \phi)}{\left(  \lambda+1\right)  \lambda^{2}%
}+\frac{2\left(  \lambda^{2}-\lambda-1\right)  \left(  \lambda+1\right)
h(\left(  \lambda-1\right)  \phi)}{\left(  \lambda-1\right)  \lambda^{2}}%
\end{align}
where, for simplifying the notation, by $h(\phi)$ we indicate an arbitrary
function of the dilaton. With these notations we can now write down the
dilaton potential and the remaining unknown in the metric as%

\begin{align}
V(\phi)  &  =\alpha_{1}W(\cosh(\phi))+\alpha_{2}W(\sinh(\phi))+4\alpha
_{0}(2\phi+\phi\cosh(\phi)-3\sinh(\phi))\nonumber\\
&  +\left(  \frac{4\alpha_{2}}{\lambda^{2}\left(  \lambda^{2}-1\right)
}\right)  \left(  \cosh(\phi)+2\right)
\end{align}

\begin{align}
f_{\lambda}(x)  &  =\frac{\eta^{2}(x-1)^{2}k}{x}+(x-\frac{1}{x}-2\frac{\ln
(x)}{x})\alpha_{0}+\frac{\eta^{2}(\lambda\left(  x-1\right)  \left(
x(\lambda+1)+1-\lambda\right)  +2x\left(  1-x^{\lambda}\right)  )Q^{2}%
}{(\lambda^{2}-1)x^{\lambda+1}}\nonumber\\
&  +\frac{\eta^{4}\left(  -2x^{\lambda+1}\left(  \lambda^{2}-1\right)
+\lambda x^{\lambda}\left(  \lambda+1\right)  -2x+\lambda x^{\lambda+2}\left(
\lambda-1\right)  \right)  P^{2}}{(\lambda^{2}-1)x\lambda^{2}}%
\end{align}


\section{A general family of electrically charged dilatonic black holes}

\label{SUGRA2}

Let us consider again the action (\ref{actionA}) and the dilaton expression as%
\begin{equation}
\phi=\sqrt{\nu^{2}-1}\ln(x)\,\,\,\,\,,\,\,\,\,\,\,\qquad l_{\nu}=\frac
{1}{\sqrt{\nu^{2}-1}}%
\end{equation}
A general class of solutions can be found when
\begin{equation}
\lambda=\left(  \frac{\nu+1}{\nu-1}\right)  ^{1/2}%
\end{equation}
for which the metric and the gauge fields are
\begin{equation}
ds^{2}=\Omega(x)\left(  -f(x)dt^{2}+\frac{\eta^{2}dx^{2}}{f(x)}+d\Sigma
_{k}\right)  \,\,\,\,\,\,\,,\,\,\,\,\,\,\,A=-\frac{Qx^{-\nu}}{{\nu}}dt
\label{gf}%
\end{equation}
where the conformal factor and the other function in the metric are
\begin{equation}
\Omega(x)=\frac{\nu^{2}x^{\nu-1}}{\eta^{2}\left(  x^{\nu}-1\right)  ^{2}}%
\end{equation}

Using {$p=\lambda\sqrt{\nu^{2}-1}=\nu+1$ and }${P=0}${ in (\ref{general}) we
get}%
\begin{align}
f(x)  &  =\frac{\eta^{2}}{\nu}\left(  x^{2}+\frac{2x^{2-\nu}}{\nu-2}-\frac
{\nu}{\nu-2}\right)  +f_{1}\left(  \frac{x^{\nu+2}}{\nu+2}-x^{2}%
+\frac{x^{2-\nu}}{2-\nu}+\frac{\nu^{2}}{\nu^{2}-4}\right) \nonumber\\
&  -\frac{Q^{2}\eta^{2}}{\nu^{3}}\left(  \frac{x^{2}}{2}+\frac{x^{2-2\nu}%
}{2-2\nu}-2\frac{x^{2-\nu}}{2-\nu}-\frac{\nu^{2}}{\left(  2-\nu\right)
\left(  2-2\nu\right)  }\right)
\end{align}
Now, we set $f_{1}=\frac{\tilde{\alpha}}{\nu^{2}}-\frac{Q^{2}\eta^{2}\left(
\nu+2\right)  }{2\nu^{3}\left(  \nu-1\right)  }+\frac{\eta^{2}\left(
\nu+2\right)  }{\nu^{2}}$ to get%

\begin{equation}
f(x)=\frac{\eta^{2}x^{2}(x^{\nu}-1)^{2}}{x^{\nu}\nu^{2}}+\frac{\tilde{\alpha}%
}{\nu^{2}}\left(  \frac{x^{\nu+2}}{\nu+2}-x^{2}+\frac{x^{2-\nu}}{2-\nu}%
+\frac{\nu^{2}}{\nu^{2}-4}\right)  -\frac{Q^{2}\eta^{2}x^{2-2\nu}}{2\nu
^{3}\left(  \nu-1\right)  }\left(  x^{\nu}-1\right)  ^{3}%
\end{equation}
The dilaton potential is%
\begin{equation}
V(\phi)=\frac{2\tilde{\alpha}}{\nu^{2}}\left(  \frac{\nu-1}{\nu+2}%
\sinh(\left(  1+\nu\right)  \phi l_{\nu})+\frac{\nu+1}{\nu-2}\sinh(\left(
1-\nu\right)  \phi l_{\nu})+4\frac{\nu^{2}-1}{\nu^{2}-4}\sinh\left(  \phi
l_{\nu}\right)  \right)  .
\end{equation}

In the limit $\nu=2$ we obtain the solution presented in Section $2.1.2$,
which is a smooth deformation of the electrically charged Kaluza-Klein black hole.



\begin{thebibliography}{99}                                                                                               %


\bibitem {Israel:1967wq}W.~Israel, ``{Event horizons in static vacuum
space-times},'' \emph{Phys.Rev.} \textbf{164} (1967) 1776--1779$\!\!$;\newline
B.~Carter, ``{Axisymmetric Black Hole Has Only Two Degrees of Freedom},''
\emph{ Phys.Rev.Lett.} \textbf{26} (1971) 331--333$\!\!$;\newline R.~H. Price,
``{Nonspherical perturbations of relativistic gravitational collapse. 1.
Scalar and gravitational perturbations},'' \emph{Phys.Rev.} \textbf{ D5}
(1972) 2419--2438$\!\!$;\newline R.~H. Price, ``{Nonspherical Perturbations of
Relativistic Gravitational Collapse. II. Integer-Spin, Zero-Rest-Mass
Fields},'' \emph{Phys.Rev.} \textbf{ D5} (1972) 2439--2454$\!\!$;\newline
S.~W. Hawking, ``{Breakdown of Predictability in Gravitational Collapse},''
\emph{Phys. Rev.} \textbf{D14} (1976) 2460--2473$\!\!$;\newline
J.~D.~Bekenstein, \textit{`No Hair': Twenty--five Years After}, chapter in
\textit{ Proceedings of the Second International Andrei D. Sakharov Conference
in Physics\/}, edited by I.~M.~Dremin and A.~M.~Semikhatov (World Scientific,
Singapore, 1997).



\bibitem {Bekenstein:1995un}J.~D.~Bekenstein, ``Novel 'no scalar hair' theorem
for black holes,'' Phys.\ Rev.\ D \textbf{51}, 6608 (1995).




\bibitem {Sudarsky:1995zg}D.~Sudarsky, ``A Simple proof of a no hair theorem
in Einstein Higgs theory,'' Class.\ Quant.\ Grav.\ \textbf{12}, 579
(1995);\newline D.~Sudarsky and T.~Zannias, ``Spherical black holes cannot
support scalar hair,'' Phys.\ Rev.\ D \textbf{58} (1998) 087502
[gr-qc/9712083].


\bibitem {Hertog:2006rr}T.~Hertog, ``Towards a Novel no-hair Theorem for Black
Holes,'' Phys.\ Rev.\ D \textbf{74}, 084008 (2006) [gr-qc/0608075].


\bibitem {Torii:2001pg}T.~Torii, K.~Maeda and M.~Narita, ``Scalar hair on the
black hole in asymptotically anti-de Sitter space-time,'' Phys.\ Rev.\ D
\textbf{64}, 044007 (2001);\newline C.~Martinez, R.~Troncoso and J.~Zanelli,
``Exact black hole solution with a minimally coupled scalar field,''
Phys.\ Rev.\ D \textbf{70}, 084035 (2004) [hep-th/0406111];\newline
E.~Winstanley, ``On the existence of conformally coupled scalar field hair for
black holes in (anti-)de Sitter space,'' Found.\ Phys.\ \textbf{33}, 111
(2003) [gr-qc/0205092];\newline A.~Anabalon and J.~Oliva, ``Exact Hairy Black
Holes and their Modification to the Universal Law of Gravitation,''
Phys.\ Rev.\ D \textbf{86}, 107501 (2012) [arXiv:1205.6012 [gr-qc]];\newline
J.~Aparicio, D.~Grumiller, E.~Lopez, I.~Papadimitriou and S.~Stricker,
``Bootstrapping gravity solutions,'' JHEP \textbf{1305}, 128 (2013)
[arXiv:1212.3609 [hep-th]];\newline W.~Xu and L.~Zhao, ``Charged black hole
with a scalar hair in (2+1) dimensions,'' arXiv:1305.5446 [gr-qc];\newline
L.~Zhao, W.~Xu and B.~Zhu, ``Novel rotating hairy black hole in
(2+1)-dimensions,'' arXiv:1305.6001 [gr-qc];\newline F.~Correa, A.~Faundez and
C.~Martinez, ``Rotating hairy black hole and its microscopic entropy in three
spacetime dimensions,'' Phys.\ Rev.\ D \textbf{87}, 027502 (2013)
[arXiv:1211.4878 [hep-th]];\newline A.~Anabalon, ``Exact Hairy Black Holes,''
arXiv:1211.2765 [gr-qc];\newline A.~Anabalon, F.~Canfora, A.~Giacomini and
J.~Oliva,``Black Holes with Primary Hair in gauged N=8 Supergravity,'' JHEP
\textbf{1206} (2012) 010 [arXiv:1203.6627 [hep-th]];\newline A.~Anabalon and
A.~Cisterna, ``Asymptotically (anti) de Sitter Black Holes and Wormholes with
a Self Interacting Scalar Field in Four Dimensions,'' Phys.\ Rev.\ D
\textbf{85} (2012) 084035 [arXiv:1201.2008 [hep-th]];\newline A.~Anabalon and
H.~Maeda, ``New Charged Black Holes with Conformal Scalar Hair,''
Phys.\ Rev.\ D \textbf{81} (2010) 041501 [arXiv:0907.0219 [hep-th]];\newline%
``Conformally coupled scalar black holes admit a flat horizon due to
axionic charge,'' JHEP \textbf{1209} (2012) 008 [arXiv:1205.4025
[hep-th]];\\
  H.~Lu, Y.~Pang and C.~N.~Pope,
  ``AdS Dyonic Black Hole and its Thermodynamics,''
  arXiv:1307.6243 [hep-th].

\bibitem {Sudarsky:2002mk}D.~Sudarsky and J.~A.~Gonzalez, ``On black hole
scalar hair in asymptotically anti-de Sitter space-times,'' Phys.\ Rev.\ D
\textbf{67}, 024038 (2003) [gr-qc/0207069];\newline U.~Nucamendi and
M.~Salgado, ``Scalar hairy black holes and solitons in asymptotically flat
space-times,'' Phys.\ Rev.\ D \textbf{68}, 044026 (2003) [gr-qc/0301062].

\bibitem {Anabalon:2012ta}A.~Anabalon, ``Exact Black Holes and Universality in
the Backreaction of non-linear Sigma Models with a potential in (A)dS4,''
arXiv:1204.2720 [hep-th];\newline A.~Acena, A.~Anabalon and D.~Astefanesei,
``Exact hairy black brane solutions in $AdS_{5}$ and holographic RG flows,''
arXiv:1211.6126 [hep-th].



\bibitem {Nunez:1996xv}D.~Nunez, H.~Quevedo and D.~Sudarsky, ``Black holes
have no short hair,'' Phys.\ Rev.\ Lett.\ \textbf{76}, 571 (1996)
[gr-qc/9601020].


\bibitem {noi}
    A.~Anabalon and D.~Astefanesei,
  ``On attractor mechanism of $AdS_{4}$ black holes,''
  arXiv:1309.5863 [hep-th].

\bibitem {Salvio:2012at}A.~Salvio, ``Holographic Superfluids and
Superconductors in Dilaton-Gravity,'' JHEP \textbf{1209}, 134 (2012)
[arXiv:1207.3800 [hep-th]];\newline A.~Salvio, ``Transitions in Dilaton
Holography with Global or Local Symmetries,'' JHEP \textbf{1303}, 136 (2013)
[arXiv:1302.4898 [hep-th]].



\bibitem {Gibbons:1987ps}G.~W.~Gibbons and K.~-i.~Maeda, ``Black Holes and
Membranes in Higher Dimensional Theories with Dilaton Fields,''
Nucl.\ Phys.\ B \textbf{298}, 741 (1988).


\bibitem {Ferrara:1995ih}S.~Ferrara, R.~Kallosh, and A.~Strominger,
\textit{N=2 extremal black holes}, \emph{Phys. Rev.} \textbf{D52} (1995)
5412--5416, [{\texttt{hep-th/9508072}}];\newline A.~Strominger,
\textit{Macroscopic entropy of $n=2$ extremal black holes}, \emph{ Phys.
Lett.} \textbf{B383} (1996) 39--43, [{\texttt{hep-th/9602111}}];\newline
S.~Ferrara and R.~Kallosh, \textit{Supersymmetry and attractors}, \emph{Phys.
Rev.} \textbf{D54} (1996) 1514--1524, [{\texttt{hep-th/9602136}}];\newline
S.~Ferrara and R.~Kallosh, \textit{Universality of supersymmetric attractors},
\emph{Phys. Rev.} \textbf{D54} (1996) 1525--1534, [{\texttt{hep-th/9603090}}].



\bibitem {Astefanesei:2007vh}D.~Astefanesei, H.~Nastase, H.~Yavartanoo and
S.~Yun, ``Moduli flow and non-supersymmetric AdS attractors,'' JHEP
\textbf{0804}, 074 (2008) [arXiv:0711.0036 [hep-th]];\newline D.~Astefanesei,
N.~Banerjee and S.~Dutta, ``(Un)attractor black holes in higher derivative AdS
gravity,'' JHEP \textbf{0811}, 070 (2008) [arXiv:0806.1334 [hep-th]].

\bibitem{Garfinkle:1990qj}
  D.~Garfinkle, G.~T.~Horowitz and A.~Strominger,
  ``Charged black holes in string theory,''
  Phys.\ Rev.\ D {\bf 43} (1991) 3140
   [Erratum-ibid.\ D {\bf 45} (1992) 3888].
\bibitem{Gregory:1992kr}
  R.~Gregory and J.~A.~Harvey,
  ``Black holes with a massive dilaton,''
  Phys.\ Rev.\ D {\bf 47} (1993) 2411
  [hep-th/9209070].

\bibitem{Charmousis:2009xr}
  C.~Charmousis, B.~Gouteraux and J.~Soda,
  ``Einstein-Maxwell-Dilaton theories with a Liouville potential,''
  Phys.\ Rev.\ D {\bf 80} (2009) 024028
  [arXiv:0905.3337 [gr-qc]].


\bibitem {Dobiasch:1981vh}P.~Dobiasch and D.~Maison, ``Stationary, Spherically
Symmetric Solutions Of Jordan's Unified Theory Of Gravity And
Electromagnetism,'' Gen.\ Rel.\ Grav.\ \textbf{14}, 231 (1982).




\bibitem {Gibbons:1985ac}G.~W.~Gibbons and D.~L.~Wiltshire, ``Black Holes in
Kaluza-Klein Theory,'' Annals Phys.\ \textbf{167}, 201 (1986)
[Erratum-ibid.\ \textbf{176}, 393 (1987)].




\bibitem {Astefanesei:2006sy}D.~Astefanesei, K.~Goldstein and S.~Mahapatra,
``Moduli and (un)attractor black hole thermodynamics,''
Gen.\ Rel.\ Grav.\ \textbf{40}, 2069 (2008) [hep-th/0611140].




\bibitem {Kallosh:1992ii}R.~Kallosh, A.~D.~Linde, T.~Ortin, A.~W.~Peet and
A.~Van Proeyen, ``Supersymmetry as a cosmic censor,'' Phys.\ Rev.\ D
\textbf{46}, 5278 (1992) [hep-th/9205027].




\bibitem {Gibbons:1996af}G.~W.~Gibbons, R.~Kallosh and B.~Kol, ``Moduli,
scalar charges, and the first law of black hole thermodynamics,''
Phys.\ Rev.\ Lett.\ \textbf{77}, 4992 (1996) [hep-th/9607108].




\bibitem {Dabholkar:2006tb}A.~Dabholkar, A.~Sen and S.~P.~Trivedi, ``Black
hole microstates and attractor without supersymmetry,'' JHEP \textbf{0701},
096 (2007) [hep-th/0611143].




\bibitem {Sen:2007qy}A.~Sen, ``Black Hole Entropy Function, Attractors and
Precision Counting of Microstates,'' Gen.\ Rel.\ Grav.\ \textbf{40}, 2249
(2008) [arXiv:0708.1270 [hep-th]].




\bibitem {Astefanesei:2010dk}D.~Astefanesei, N.~Banerjee and S.~Dutta,
``Moduli and electromagnetic black brane holography,'' JHEP \textbf{1102}, 021
(2011) [arXiv:1008.3852 [hep-th]];\newline D.~Astefanesei, N.~Banerjee and
S.~Dutta, ``Near horizon data and physical charges of extremal AdS black
holes,'' Nucl.\ Phys.\ B \textbf{853}, 63 (2011) [arXiv:1104.4121 [hep-th]].




\bibitem {Kunduri:2007vf}H.~K.~Kunduri, J.~Lucietti and H.~S.~Reall,
``Near-horizon symmetries of extremal black holes,''
Class.\ Quant.\ Grav.\ \textbf{24}, 4169 (2007) [arXiv:0705.4214
[hep-th]];\newline D.~Astefanesei and H.~Yavartanoo, ``Stationary black holes
and attractor mechanism,'' Nucl.\ Phys.\ B \textbf{794}, 13 (2008)
[arXiv:0706.1847 [hep-th]].


\bibitem {Goldstein:2005hq}K.~Goldstein, N.~Iizuka, R.~P. Jena, and S.~P.
Trivedi, \textit{Non-supersymmetric attractors}, \emph{Phys. Rev.}
\textbf{D72} (2005) 124021, [{\texttt{hep-th/0507096}}].

\bibitem {Sen:2005wa}A.~Sen, \textit{Black hole entropy function and the
attractor mechanism in higher derivative gravity}, \emph{JHEP} \textbf{09}
(2005) 038, [{\texttt{hep-th/0506177}}];\newline A.~Sen, \textit{Entropy
function for heterotic black holes}, \emph{JHEP} \textbf{03} (2006) 008,
[{\texttt{hep-th/0508042}}].

\bibitem {Astefanesei:2006dd}D.~Astefanesei, K.~Goldstein, R.~P. Jena, A.~Sen,
and S.~P. Trivedi, \textit{ Rotating attractors}, \emph{JHEP} \textbf{10}
(2006) 058, [{\texttt{hep-th/0606244}}].



\bibitem {Allahverdi:2006we}R.~Allahverdi, K.~Enqvist, J.~Garcia-Bellido,
A.~Jokinen and A.~Mazumdar, ``MSSM flat direction inflation: Slow roll,
stability, fine tunning and reheating,'' JCAP \textbf{0706}, 019 (2007)
[hep-ph/0610134].




\bibitem {Barcelo:2000zf}E.~E.~Flanagan and R.~M.~Wald, ``Does back reaction
enforce the averaged null energy condition in semiclassical gravity?,''
Phys.\ Rev.\ D \textbf{54}, 6233 (1996) [gr-qc/9602052];\newline C.~Barcelo
and M.~Visser, ``Scalar fields, energy conditions, and traversable
wormholes,'' Class.\ Quant.\ Grav.\ \textbf{17}, 3843 (2000)
[gr-qc/0003025];\newline M.~Salgado, D.~Sudarsky and U.~Nucamendi, ``The
Violation of the weak energy condition, is it generic of spontaneous
scalarization?,'' Phys.\ Rev.\ D \textbf{70}, 084027 (2004) [gr-qc/0402126].


\bibitem {Astefanesei:2005ad}D.~Astefanesei and E.~Radu, ``Quasilocal
formalism and black ring thermodynamics,'' Phys.\ Rev.\ D \textbf{73}, 044014
(2006) [hep-th/0509144];\newline R.~B.~Mann and D.~Marolf, ``Holographic
renormalization of asymptotically flat spacetimes,''
Class.\ Quant.\ Grav.\ \textbf{23}, 2927 (2006) [hep-th/0511096];\newline
D.~Astefanesei, R.~B.~Mann, M.~J.~Rodriguez and C.~Stelea, ``Quasilocal
formalism and thermodynamics of asymptotically flat black objects,''
Class.\ Quant.\ Grav.\ \textbf{27}, 165004 (2010) [arXiv:0909.3852
[hep-th]];\newline D.~Astefanesei, M.~J.~Rodriguez and S.~Theisen,
``Thermodynamic instability of doubly spinning black objects,'' JHEP
\textbf{1008}, 046 (2010) [arXiv:1003.2421 [hep-th]].




\bibitem {Leblond:2002ns}F.~Leblond, D.~Marolf and R.~C.~Myers, ``Tall tales
from de Sitter space 1: Renormalization group flows,'' JHEP \textbf{0206}, 052
(2002) [hep-th/0202094];\newline D.~Astefanesei, R.~B.~Mann and E.~Radu,
``Reissner-Nordstrom-de Sitter black hole, planar coordinates and dS / CFT,''
JHEP \textbf{0401}, 029 (2004) [hep-th/0310273].

\end{thebibliography}
\end{document}